\documentclass[aps,twocolumn]{revtex4}
\usepackage{graphicx}
\usepackage{amsmath}
\usepackage{amssymb}
\usepackage{microtype}
\usepackage[usenames,dvipsnames,svgnames,table]{xcolor}

\begin{document}

\title{Synthetic dimensions and spin-orbit coupling with an optical clock transition}

\author{L. F. Livi$^{1,5}$, G. Cappellini$^{2,5}$, M. Diem$^{3,4}$, L. Franchi$^2$, C. Clivati$^4$, M.~Frittelli$^4$, F. Levi$^4$, D. Calonico$^4$, J. Catani$^{5,1,6}$, M. Inguscio$^{2,1,5}$, L. Fallani$^{2,1,6}$}

\affiliation{
$^1$LENS European Laboratory for Nonlinear Spectroscopy, Firenze, Italy\\
$^2$Department of Physics and Astronomy, University of Florence, Italy\\
$^3$ILP Institut f\"ur Laserphysik, Universit\"at Hamburg, Germany\\
$^4$INRIM Istituto Nazionale di Ricerca Metrologica, Torino, Italy\\
$^5$INO-CNR Istituto Nazionale di Ottica del CNR, Sezione di Sesto Fiorentino, Italy\\
$^6$INFN Istituto Nazionale di Fisica Nucleare, Sezione di Firenze, Italy
}

\begin{abstract}
We demonstrate a novel way of synthesizing spin-orbit interactions in ultracold quantum gases, based on a single-photon optical clock transition coupling two long-lived electronic states of two-electron $^{173}$Yb atoms. By mapping the electronic states onto effective sites along a synthetic ``electronic'' dimension, we have engineered synthetic fermionic ladders with tunable magnetic fluxes. We have detected the spin-orbit coupling with fiber-link-enhanced clock spectroscopy and directly measured the emergence of chiral edge currents, probing them as a function of the magnetic field flux. These results open new directions for the investigation of topological states of matter with ultracold atomic gases.
\end{abstract}

\maketitle

\noindent
Ultracold atoms are emerging as a very versatile platform for the investigation of topological states of matter \cite{goldman16}, thanks to the possibility of using laser light to synthesize artificial gauge fields \cite{dalibard11,goldman14} and to engineer lattices with topological band structures \cite{struck12,aidelsburger13,miyake13,jotzu14,duca15}. A prime element for the emergence of nontrivial topological properties is the presence of spin-orbit coupling (SOC) \cite{galitski13, zhai15}, locking the spin of the particles to their motion. This interaction was first synthesized in cold atomic gases by using two-photon Raman transitions \cite{lin11}, which couple two different hyperfine spin states of the electronic ground-state manifold with a transfer of momentum. The coupling between spin states also enables a new powerful tool for engineering topological states of matter, which relies on the ``synthetic dimension'' concept \cite{boada12,celi14}. According to this approach, the internal states of an atom are treated as effective sites along a synthetic lattice dimension, and coherent coupling between them is interpreted in terms of an effective tunnelling between the sites. This idea has been recently realized in Refs. \cite{mancini15,stuhl15}, where synthetic flux ladders have been implemented by using the atomic spin degree of freedom, and has allowed the first observation of chiral edge states in ultracold atomic systems. Its extension has inspired several theoretical proposals, opening the way e.g. to the observation of new quantum states \cite{barbarino15,budich15}, to the detection of fractional charge pumping \cite{zeng15,taddia16}, or to the possible observation of the four-dimensional quantum Hall effect \cite{price15}.

\begin{figure}[t!]
\begin{center}
\includegraphics[width=\columnwidth]{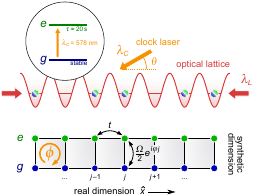}
\end{center}
\caption{Sketch of the experimental setup. Ultracold $^{173}$Yb fermions are trapped in one-dimensional chains by an optical lattice at wavelength $\lambda_L$. An ultranarrow clock laser with wavelength $\lambda_C$ drives the single-photon transition between the ground state $g=$ $^1$S$_0$ and the long-lived electronic state $e=$ $^3$P$_0$. The momentum transfer $\delta k=2\pi \cos \theta/\lambda_C$ in the atom-laser interaction results in a locking between internal state (interpreted as an effective pseudospin) and atomic momentum. The electronic state can be also treated as an effective ``synthetic dimension'' made by two sites connected with a coherent tunnelling, resulting in a two-leg fermionic ladder pierced by a synthetic magnetic field flux $\phi=\pi \delta k / k_L$ per plaquette. The flux can be tuned by adjusting the angle $\theta$ between the clock laser and the lattice.}
\end{figure}

\begin{figure*}[t!]
\begin{center}
\includegraphics[width=1.68\columnwidth]{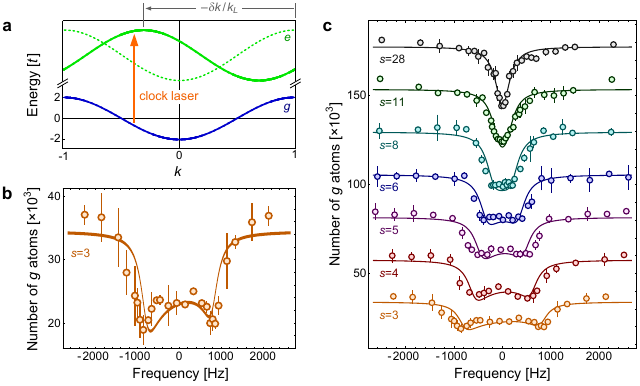}
\end{center}
\caption{Spectroscopic observation of spin-orbit coupling. {\bf a}. Sketch of the energy bands induced by the optical lattice for the $g$ (blue) and $e$ (green dashed) states. Shifting the $e$ dispersion relation by the SOC momentum $-\delta k$ (green solid) allows us to represent the clock transition between the two states as a "vertical" process with a momentum-dependent energy. {\bf b}. Spectrum of the clock transition for spin-polarized fermions in an optical lattice with $s=3$, $t=2\pi \times 220$ Hz and $\delta k=1.31 k_L$. The plot shows the number of atoms remaining in the $g$ state after a 800-ms long interrogation time. The horizontal axis shows the detuning with respect to the center of the spectrum. The shape of the spectrum, featuring a broadening and an enhanced response at the edges, is a spectroscopic consequence of spin-orbit coupling. Points are averages over multiple measurements and error bars are standard deviations. {\bf c}. Spectra of the clock transition for different lattice depths. Datasets with $s \geq 4$ have been offset vertically for the sake of presentation. The curves are the result of a single-particle theoretical model, with vertical amplitude and offset as only fit parameters. All the experimental spectra have been measured shining the clock laser light along the lattice direction ($\theta=0$) and using only an additional transverse (vertical) optical lattice to hold the atoms against gravity.}
\end{figure*}

In this Letter we demonstrate that spin-orbit coupling and synthetic dimensions can be efficiently implemented by exploiting different degrees of freedom, specifically, the long-lived atomic {\it electronic} state of alkaline-earth(-like) atoms. By using the technology developed in the context of optical atomic clocks, we induce a coherent coupling between the ground state $g=$ $^1$S$_0$ and the metastable excited state $e=$ $^3$P$_0$ (lifetime $\sim$20 s) of ultracold $^{173}$Yb atoms.
Since the two states are separated by an optical energy, it is possible to have a sizable transfer of momentum with a single-photon transition, which induces a non-negligible spin-orbit interaction when the two states are treated as spin projection states of an effective $J=1/2$ spin, as suggested in Ref. \cite{wall16}. Hereafter, we will use the term ``spin'' to refer to this pseudospin degree of freedom. An advantage of this pure two-level scheme over the Raman scheme employed so far is that it avoids the detrimental effect of near-resonant intermediate states that have been observed to cause strong heating in alkali fermions \cite{wang12,cheuk12} hampering the observation of true many-body effects \cite{nota1}. Moreover, the implementation of this strategy relies on an simpler experimental configuration consisting of a single laser beam, which simplifies the alignment procedure and does not suffer from interferometric instabilities. The effectiveness of this stategy, in terms of easier experimental implementation, allows us to study chiral edge currents in synthetic fermionic ladders with a fully tunable synthetic magnetic flux.

In the experiment, we trap ultracold Fermi gases of $^{173}$Yb in a one-dimensional optical lattice potential $V(x)=s E_R \cos^2 k_L x$ generated by laser light at the ``magic'' wavelength $\lambda_L=2\pi/k_L=759$ nm, as sketched in Fig. 1. The depth of the optical lattice $s$ (measured in units of the recoil energy $E_R=\hbar^2 k_L^2 / 2m$, where $m$ is the atomic mass) is the same for both the states $g$ and $e$ and determines the tunnelling energy $t$ between next-neighboring lattice sites. An additional 2D transverse lattice, not depicted in Fig. 1, freezes the atomic motion along the two orthogonal directions, producing an array of one-dimensional independent fermionic wires. In order to engineer the spin-orbit coupling, we drive the $g-e$ transition using $\pi$-polarized light of an ultranarrow $\lambda_C=578$ nm clock laser, frequency-stabilized to an Ultra Low Expansion glass cavity, as described in Ref. \cite{cappellini15,pizzocaro2012}. The short-term linewidth of the clock laser is of the order of 30 Hz on a 1 s timescale, as evidenced by spectroscopic measurements. In order to cancel the residual long-term drift of the clock laser, we discipline it to a stable optical frequency reference generated at the Italian National Institute for Metrological Research (INRiM) in Turin and delivered to our laboratory in Florence trough an optical fiber link infrastructure \cite{clivati16,calonico14}. The angle $\theta$ between the clock laser and the optical lattice (see Fig. 1) can be changed, resulting in an effective, tunable momentum transfer $\delta k=2\pi \cos \theta / \lambda_C$ along the direction of the atomic chains. 

In a first set of experiments we demonstrate the capability of the optical clock transition to induce the momentum transfer required for the generation of spin-orbit coupling. Fig. 2a shows a diagram with the energy spectrum of the lowest optical lattice band for $g$ (blue line) and $e$ (green dotted line) states as a function of the lattice momentum $k$, defined in units of $k_L$. The lattice dispersions are the same for both the $g$ (blue line) and $e$ (green dotted line) states, since the lattice is operated at the magic wavelength. Because of the conservation of momentum in the atom-light interaction, the clock transition connects state $|g,k\rangle$ with state $|e,k+\delta k/k_L\rangle$, ensuring spin-momentum locking. By performing a standard gauge transformation, it is possible to sketch the transition as a momentum-preserving process (vertical arrow) between the $g$ band and a momentum-shifted $e$ band (green thick line), which enlights the dependence of the transition energy on the momentum state $k$. Fig. 2b shows a spectrum of the clock transition for a band insulator of nuclear-spin-polarized $^{173}$Yb atoms trapped in a lattice with $s=3$, corresponding to a tunnelling constant $t=2\pi \times 220$ Hz. The spectrum shows an enhanced response at the edges, that is related to a divergent density of states induced by the van Hove singularities of the optical lattice \cite{ashcroft76}. The width of the spectrum is related to the momentum transfer and to the tunnelling strength, being $8t \, |\sin (\pi \delta k / 2 k_L)|$ for a fully occupied first lattice band. The observation of this peculiar lineshape is a first spectroscopic signature of spin-orbit coupling, as pointed out in Ref. \cite{wall16}. Fig. 2c shows a collection of spectra for different lattice depths, from $s=28$ to $s=3$, illustrating the crossover between clock spectroscopy in the Lamb-Dicke regime at large $s$ (where the lattice bands are flat) to momentum-selective excitations at small $s$. The curves in Fig. 2b and Fig. 2c are predictions of a single-particle theoretical model in which the only adjusted parameters are a vertical offset and a vertical scaling factor. The theoretical spectra have been calculated by assuming a uniform population of the lowest lattice band (as verified experimentally from band-mapping measurements) and convolved with the experimental spectroscopic resolution function. The latter can be approximated with a power-broadened Lorentzian line profile with a half width at half maximum of 170 Hz (for the data in Fig. 2), derived from a fit of the measured spectrum at $s=28$. The agreement with the calculation is quite good, evidencing the capability of the optical clock excitation to address the energy band in a momentum- and energy-selective way.

\begin{figure}[t!]
\begin{center}
\includegraphics[width=0.92\columnwidth]{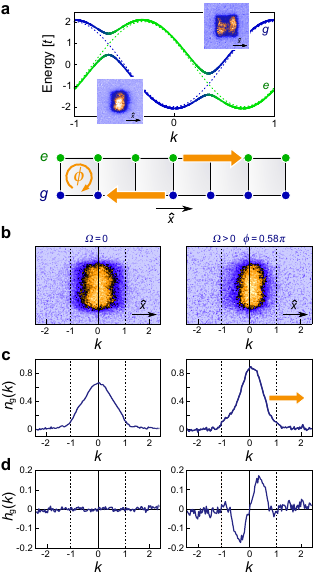}
\end{center}
\caption{Chiral currents in the synthetic ladder. {\bf a}. Sketch of the dressed (thick, solid) and bare energy bands (thin, dashed) for $\phi=1.31 \pi$ and $\Omega =t$. The color of the thick lines reflects the state composition of the dressed bands ($e$: green, $g$: blue): the different composition as a function of $k$ can be interpreted as chiral currents flowing in opposite directions along the two legs of the synthetic ladder. The insets show time-of-flight images of the atoms in the $g$ state for each of the two dressed bands after adiabatic loading (for $\phi=1.31 \pi$, $t=2\pi \times 138$ Hz and $\Omega=2\pi \times 590$ Hz), evidencing complementary momentum distributions. {\bf b}. Time-of-flight images of the $g$ atoms for decoupled legs ($\Omega=0$, left) and for coupled legs after adiabatic loading ($\Omega=2\pi \times 590$ Hz, $\phi=0.58\pi$, $t=2\pi \times 138$ Hz, right). {\bf c}. Integrated lattice momentum distribution $n_g(k)$ along the ladder direction for the data reported in panel b. {\bf d}. Asymmetry function $h_g(k)=n_g(k)-n_g(-k)$, evidencing the chiral current in the presence of the flux.}
\end{figure}

In a second set of experiments, we exploit the spin-orbit coupling induced by the clock laser to engineer synthetic flux ladders following the ``synthetic dimension'' approach \cite{celi14}. In this picture, the coherent coupling between $g$ and $e$ can be interpreted as the realization of a two-leg flux ladder geometry, with the two legs corresponding to the states $g$ and $e$, a complex tunnelling along the rungs with amplitude $\Omega/2$  ($\Omega$ being the Rabi frequency associated to the clock excitation), and a synthetic magnetic field flux $\phi = \pi \delta k / k_L = \delta k d$ per plaquette (, where $d=\lambda_L/2$ is the lattice spacing), as also sketched in Fig. 1. At the single-particle level, the system is  described by the Harper-Hofstadter ladder Hamiltonian
\begin{equation}
H = -\hbar \left( t\sum_{j,\alpha} c^{\dagger}_{j, \alpha}c_{j+1, \alpha}+ \frac{\Omega}{2} \sum_{j} e^{i \phi j}c^\dagger_{j,e}c_{j,g}\right) +\textrm{h.c.} \; ,
\nonumber
\end{equation}
where $c^{\dagger}_{j, \alpha}$ ($c_{j, \alpha}$) are fermionic creation (annihilation) operators on the site ($j,\alpha$) in the real ($j$) and synthetic ($\alpha = e,\,g$) dimension. Fig. 3a shows the dressed energy bands of this Hamiltonian (after the gauge transformation described above) as a function of the lattice momentum along $\hat{x}$, with a gap opening in correspondence of the crossing between the two bare-state energy curves. The color of the dressed bands indicates the $g-e$ mixing of the different momentum states, evidencing the spin-momentum locking. The difference in momentum-resolved state composition between lower and higher dressed bands can be clearly seen in the false-color time-of-flight images of the $g$ atoms in Fig. 3a, obtained with the technique outlined below.

In the synthetic dimension picture, spin-momentum locking corresponds to chiral edge currents travelling in opposite direction along the two legs of the ladder, as sketched in Fig. 3a. In order to detect these currents, we prepare the system in an equilibrium state by performing an adiabatic sweep of the clock laser frequency that loads the atoms, initially in the $g$ state, into the lowest-energy (stationary) dressed state. Fig. 3b shows false-color absorption images of the atoms in the $g$ state after a sudden switch off of the clock laser and a band-mapping procedure to measure the lattice momentum distribution \cite{mancini15}, for two different cases: without the clock laser ($\Omega=0$, $t=2\pi \times 138$ Hz, left) and with the clock laser generating the flux ladder ($\Omega=2\pi \times 590$ Hz, $\phi=0.58 \pi$, $t=2\pi \times 138$ Hz, right). Fig. 3c shows the momentum distribution $n_g(k)$ along the lower leg of the ladder (normalized in order to have $\int n_g(k) \, dk = 1$, after integration along the transverse directions), evidencing a clear asymmetry towards positive momenta when the clock laser induces a nonzero flux through the ladder plaquettes. This is more evident in Fig. 3d, where we plot the asymmetry function $h_g(k)=n_g(k)-n_g(-k)$. The chirality of the atomic motion can be quantified by introducing the momentum-integrated quantity $J=\int_{0}^1 dk \, h_g(k)$, as in Ref.  \cite{mancini15}.

\begin{figure}[t]
\begin{center}
\includegraphics[width=\columnwidth]{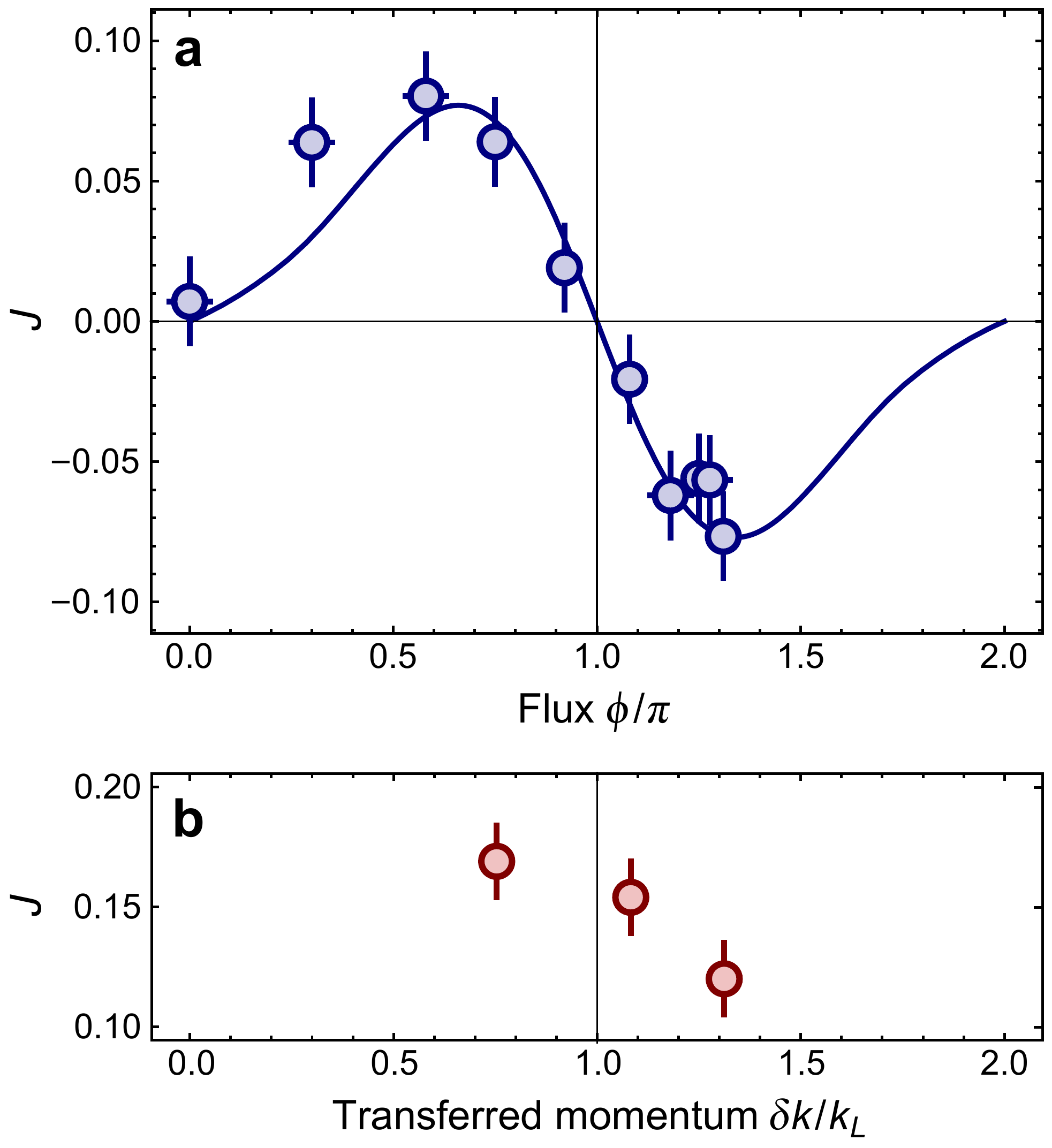}
\end{center}
\caption{{\bf a}. Dependence of the chiral currents $J$ on the flux $\phi$. Vertical error bars are the standard deviation of a representative set of measurements, while horizontal error bars arise from the uncertainty on the angle $\theta$. The solid curve is the expectation of a theoretical single-particle model (see main text). Experimental parameters: $t=2\pi \times 138$ Hz, $\Omega = 2\pi \times 590$ Hz. {\bf b}. Chiral currents in the absence of the lattice along the direction of the ladders. In this "continuum" configuration, where a 2D unit cell is not defined, no inversion of $J$ is observed at large SOC momentum transfer $\delta k$. We note that the horizontal scales of the two plots coincide as $\phi/\pi = \delta k/k_L$.}
\end{figure}

The single-photon excitation makes it experimentally easy to probe different synthetic magnetic fluxes $\phi$ (or, equivalently, different SOC momentum transfers $\delta k$), by changing the angle $\theta$ of the clock laser with respect to the orientation of the wires. In Fig. 4a we show the measured $J$ as a function of $\phi$. The data show a strong dependence of the chirality as a function of the synthetic magnetic flux. Noticeably, our data show an inversion of the sign of the chiral current when the flux per plaquette exceeds $\pi$. At a qualitative level, the observed behavior can be understood on the basis of the symmetries of the system, since the flux per plaquette is defined modulo $2\pi$ ($J(\phi)=J(\phi+2\pi)$) and the current changes sign when the flux is inverted ($J(\phi)=-J(-\phi)$) because of its chiral nature. The experimental points are compared with the result of a single-particle calculation based on the exact diagonalization of a system of fermions in the two-leg ladder, showing a very good agreement. The theoretical analysis takes into account an effective harmonic confinement along $\hat{x}$ (as in Ref. \cite{mancini15}) and the average over the distribution of atomic wires (with inhomogeneous filling) realized in the experiment. We have used the temperature as an adjustable parameter, resulting in a best-fitting value $T=0.6t$. We note however, that the shape of the curve is robust against the fine tuning of the parameters, as it is fundamentally implied by the symmetries of the problem.

In particular, the inversion of the sign of $J$ above $\pi$ flux is a direct consequence of the discreteness of the Harper-Hofstadter Hamiltonian. By removing the optical lattice along $\hat{x}$ we can synthesize a two-leg ladder in the continuum, where a unit cell is no longer defined and the symmetries arising from the discreteness of the system no longer hold. The measured values of $J$ for this configuration are reported in Fig. 4b as a function of the transferred SOC momentum $\delta k$. In this case we do not observe any inversion of $J$ when $\delta k$ exceeds $k_L$ (corresponding to the $\pi$ flux condition in Fig. 4a).

In conclusion, we have shown a new way to engineer spin-orbit coupling and synthetic ladders with ultracold gases of two-electron fermions. Our system is highly promising for further studies of spin-orbit-coupled fermions in the degenerate regime. As an extension of this work, we mention the possibility of coupling different electronic and {\it nuclear spin} states, e.g. with circularly-polarized clock light. This would enable us to control the interactions between the two internal states by using the recently discovered orbital Feshbach resonance of $^{173}$Yb atoms \cite{zhang15,pagano15,hofer15}. This would allow the investigation of synthetic flux ladders with tunable inter-leg interactions, and could be used to investigate spin-orbit coupling in atomic $^{173}$Yb Fermi superfluids at the BEC-BCS crossover \cite{xu16,he16}, e.g. to study the possible emergence of topological superfluidity. In the synthetic dimensions approach, we envision the possibility of combining the nuclear spin synthetic dimension with the electronic synthetic dimension, realizing multi-dimensional synthetic lattice structures with nontrivial connectivities \cite{boada15}, periodic boundary conditions and engineered topological properties thanks to the control of the tunnelling phases.

This work has been supported by EU FP7 SIQS, MIUR PRIN2012 AQUASIM, INFN FISh. This project has received funding from the EMPIR programme co-financed by the Participating States and from the European Union's Horizon 2020 research and innovation programme.
We thank P. Cancio Pastor for experimental assistance and M. Mancini, G. Pagano and C. Sias for the critical reading of the manuscript. We also thank TOPTICA Photonics AG for prompt technical assistance. 

We would like to point out that, during the completion of the experimental work, we became aware of closely related results obtained by spectroscopic measurements on $^{87}$Sr nondegenerate atoms in an optical lattice clock setup \cite{kolkowitz16}.


\begin{thebibliography}{99}

\bibitem{goldman16}
N. Goldman, J. C. Budich, and P. Zoller, Nature Phys. {\bf 12}, 639 (2016).

\bibitem{dalibard11}
J. Dalibard, F. Gerbier, G. Juzeli\={u}nas, P. \"{O}hberg, Rev. Mod. Phys. {\bf 83}, 1523 (2011).

\bibitem{goldman14}
N. Goldman {\it et al.}, Rep. Prog. Phys. {\bf 77}, 126401 (2014).

\bibitem{struck12}
J. Struck {\it et al.}, Phys. Rev. Lett. {\bf 108}, 225304 (2012).

\bibitem{aidelsburger13}
M. Aidelsburger {\it et al.}, Phys. Rev. Lett. {\bf 111}, 185301 (2013).

\bibitem{miyake13}
H. Miyake {\it et al.}, Phys. Rev. Lett. {\bf 111}, 185302 (2013).

\bibitem{jotzu14}
G. Jotzu {\it et al.}, Nature {\bf 515}, 237 (2014).

\bibitem{duca15}
L. Duca {\it et al.}, Science {\bf 347}, 288 (2015). 

\bibitem{galitski13}
V. Galitski and I. B. Spielman, Nature {\bf 494}, 49 (2013).

\bibitem{zhai15}
H. Zhai, Rep. Prog. Phys. {\bf 78}, 026001 (2015).

\bibitem{lin11}
Y.-J. Lin, K. Jim{\'e}nez-Garc{\'i}a, and I. B. Spielman, Nature {\bf 471}, 83 (2011).

\bibitem{boada12}
O. Boada et al., Phys. Rev. Lett. {\bf 108}, 133001 (2012).

\bibitem{celi14}
A. Celi et al., Phys. Rev. Lett. {\bf 112}, 043001 (2014).

\bibitem{mancini15}
M. Mancini {\it et al.}, Science {\bf 349}, 1510 (2015).

\bibitem{stuhl15}
B. K. Stuhl {\it et al.}, Science {\bf 349}, 1514 (2015).

\bibitem{barbarino15}
S. Barbarino, L. Taddia, D. Rossini, L. Mazza, and R. Fazio, Nat. Comm. {\bf 6}, 8134 (2015).

\bibitem{budich15}
J. C. Budich, C. Laflamme, F. Tschirsich, S. Montangero, and P. Zoller, Phys. Rev. B {\bf 92}, 245121 (2015).

\bibitem{zeng15}
T.-S. Zeng, C. Wang, and H. Zhai, Phys. Rev. Lett. {\bf 115}, 095302 (2015).

\bibitem{taddia16}
L. Taddia {\it et al.}, preprint arXiv:1607.07842 (2016).

\bibitem{price15}
H. M. Price, O. Zilberberg, T. Ozawa, I. Carusotto, and N. Goldman, Phys. Rev. Lett. {\bf 115}, 195303 (2015).

\bibitem{wall16}
M. L. Wall {\it et al.}, Phys. Rev. Lett. {\bf 116}, 035301 (2016).

\bibitem{wang12}
P. Wang {\it et al.}, Phys. Rev. Lett. {\bf 109}, 095301 (2012).

\bibitem{cheuk12}
L. W. Cheuk {\it et al.}, Phys. Rev. Lett. {\bf 109}, 095302 (2012).

\bibitem{nota1}
Different strategies to overcome this problem have been recently implemented also with Raman transitions in non-alkali fermions \cite{burdick16,song16} or by implementing effective mappings using superlattice potentials \cite{atala14,li16}.

\bibitem{burdick16}
N. Q. Burdick, Y. Tang, and B. L. Lev, Phys. Rev. X {\bf 6}, 031022 (2016).

\bibitem{song16}
B. Song {\it et al.}, preprint arXiv:1608.00478 (2016).

\bibitem{atala14}
M. Atala {\it et al.}, Nature Phys. {\bf 10}, 588 (2014).

\bibitem{li16}
J. Li {\it et al.}, preprint arXiv:1606.03514 (2016).

\bibitem{cappellini15}
G. Cappellini {\it et al.}, Rev. Sci. Instrum. {\bf 86}, 073111 (2015).

\bibitem{pizzocaro2012}
M. Pizzocaro {\it et al.}, IEEE Trans. on UFFC, {\bf 59}, 426 (2012).

\bibitem{clivati16}
C. Clivati {\it et al.}, Opt. Express {\bf 24}, 11865 (2016).

\bibitem{calonico14}
D. Calonico {\it et al.}, Appl. Phys. B {\bf 117}, 979 (2014).

\bibitem{ashcroft76}
N. W. Ashcroft and D. N. Mermin, Solid State Physics, Saunders College Publishing, Fort Worth, TX (1976).

\bibitem{zhang15}
R. Zhang, Y. Cheng, H. Zhai, and P. Zhang, Phys. Rev. Lett. {\bf 115}, 135301 (2015).

\bibitem{pagano15}
G. Pagano {\it et al.}, Phys. Rev. Lett. {\bf 115}, 265301 (2015).

\bibitem{hofer15}
M. H{\"o}fer {\it et al.}, Phys. Rev. Lett. {\bf 115}, 265302 (2015).

\bibitem{xu16}
J. Xu {\it et al.}, preprint arXiv:1602.06513 (2016).

\bibitem{he16}
L. He, J. Wang, S.-G. Peng, X.-Ji Liu, and H. Hu, preprint arXiv:1606.00188 (2016).

\bibitem{boada15}
O. Boada {\it et al.}, New J. Phys. {\bf 17}, 045007 (2015).

\bibitem{kolkowitz16}
S. Kolkowitz {\it et al.}, preprint arXiv:1608.03854 (2016).

\end{thebibliography}
\end{document}